\documentstyle[12pt]{article}
\begin{document}

\baselineskip=.285in

\catcode`\@=11
\def\maketitle{\par
 \begingroup
 \def\thefootnote{\fnsymbol{footnote}}
 \def\@makefnmark{\hbox
 to 0pt{$^{\@thefnmark}$\hss}}
 \if@twocolumn
 \twocolumn[\@maketitle]
 \else \newpage
 \global\@topnum\z@ \@maketitle \fi\thispagestyle{empty}\@thanks
 \endgroup
 \setcounter{footnote}{0}
 \let\maketitle\relax
 \let\@maketitle\relax
 \gdef\@thanks{}\gdef\@author{}\gdef\@title{}\let\thanks\relax}
\def\@maketitle{\newpage
 \null
 \hbox to\textwidth{\hfil\hbox{\begin{tabular}{r}
 \@preprint\end{tabular}}}
 \vskip 2em \begin{center}
 {\Large\bf \@title \par} \vskip 1.5em {\normalsize \lineskip .5em
\begin{tabular}[t]{c}\@author
 \end{tabular}\par}
 \end{center}
 \par
 \vskip 1.5em}
\def\preprint#1{\gdef\@preprint{#1}}
\def\abstract{\if@twocolumn
\section*{Abstract}
\else \normalsize
\begin{center}
{\large\bf Abstract\vspace{-.5em}\vspace{0pt}}
\end{center}
\quotation
\fi}
\def\endabstract{\if@twocolumn\else\endquotation\fi}
\catcode`\@=12

\preprint{DPNU-94-39\\ hep-th/9410076}

\title{
\Large {\bf Bubble with Global Monopole} \protect\\[1mm]\  }
\author{
{\normalsize Yoonbai Kim$^{\ast}$}\\ [2mm]
{\normalsize\it Department of Physics, Nagoya University,
Nagoya 464-01, Japan}}
\date{}
\maketitle

\vspace{3mm}

\begin{center}
{\large\bf Abstract}\\[3mm]
\end{center}
\indent\indent
The first-order phase transition of $O(3)$ symmetric models is
considered in
the limit of high temperature. It is shown that this model supports a
new bubble solution where the global monopole is formed at the center
of the
bubble in addition to the ordinary $O(3)$ bubble.
Though the free energy of it is
larger than that of normal bubble, the production rate can
considerably be
large at high temperatures.\newline


\def\thepage{\protect\hspace{30mm}
\protect\raisebox{2.4ex}{ }\protect\hspace{-42.5mm}
\protect\raisebox{0ex}{\ }
\protect\hspace{-23.1mm}\protect\raisebox{-2.4ex}{ }}
\thispagestyle{headings}
\markright{\thepage}

\newpage

\pagenumbering{arabic}
\thispagestyle{plain}

\def\be{\begin{equation}}
\def\ee{\end{equation}}
\def\bea{\begin{eqnarray}}
\def\eea{\end{eqnarray}}
\def\thint{\int d^3\! x\,\,}

Since a general theory of the decay of the metastable phase was
developed
\cite{Lan,Col1}, the study of first-order
phase transitions have attracted the attention due to their possible
relevance to the physics of the early universe.
The semiclassical expression of the tunneling rate from false vacuum
to true
vacuum is given by the bubble solution of lowest (Euclidean)
action both
for the first-order phase transitions at zero temperature field
theories
\cite{Col1} and those at finite temperature \cite{Lin,KT}.
However, the above case does not include the possibility that there
exist various decay modes between two given classical vacua. In this note,
we explore such problem by examining a scalar model of internal $O(3)$
symmetry at finite temperature and
examine a new bubble solution with a global monopole.

Suppose that the model of our interest contains a series of stationary points
which support bubble solutions,
a decay probability per unit time
per unit volume is given by
\be\label{decay}
\Gamma /V=\sum_{n}A_{n}e^{-B_{n}/\hbar},
\ee
where $n$ represents the $n$-th local minimum. For each $n$, $B_{n}$ is given
by a
value of Euclidean action
for $n$-th bubble solution and $A_{n}$ is
estimated by
integrating out the fluctuations around a given $n$-th
configuration.

We choose the $O(3)$-symmetric scalar models described in terms of
an isovector field $\phi^{a}$ $(a=1,2,3)$ as the first possibility for
sample calculations. At finite temperature the Euclidean action is
\be \label{action}
S_{E}=\int^{\beta}_{0}\! dt_{E} \thint\left\{
\frac{1}{2}\left(\frac{\partial\phi^{a}}{\partial t_{E}}\right)^{2}
+\frac{1}{2}\left(\nabla\phi^{a}\right)^{2}+V(\phi^{a})
\right\},
\ee
where $\beta=\hbar /k_{B}T$.
Although our argument followed is general and does scarcely depend
on the detailed form of scalar potential if it has
the vacuum structure with both a true and a false vacuum, we will consider a
specific model, {\it i.e.} a $\phi^{6}$-potential
\be
V(\phi)=\frac{\lambda}{v^{2}}(\phi^{2}+\alpha v^{2})(\phi^{2}-v^{2})^{2}
+V_{0}
\ee
where $\phi$ denotes the amplitude of scalar
fields $\phi^{a}$ defined by $\phi=\sqrt{\phi^{a}\phi^{a}}$.
Here we consider the transition from the symmetric vacuum to the broken vacuum,
{\it i.e.} $0<\alpha<1/2$, and choose $V_{0}$ as $-\lambda\alpha v^{4}$ in
order to make the value of $S_{E}$ coinside with $B_{n}$ in Eq.(\ref{decay}).
This does not lose generality since the case of the transition from the
broken vacuum to the symmetric one ($-1<\alpha<0$) also possesses the same type
of bubble solutions if we replace $\phi$ to $v-\phi$.

It has been known that the rate of vacuum transitions at finite
temperature has relevant amount of contributions from $O(3)$-symmetric
sphaleron-type bubbles \cite{Lin,KT}. In high temperature limit
this contribution dominates and then we can neglect the dependence along
$t_{E}$-axis. Here we are interested in the regime where thermal
fluctuations are much larger than quantum fluctuations and address the problem
for different kind of finite temperature
bubbles which connect global $O(3)$ internal symmetry to that of
spatial rotation. We now ask for a solution of the field equations
that is time-independent and spherically symmetric, apart from the
angle dependence due to the mapping between the $(\theta,\varphi)$ angles
in space and those of isovector space $\hat{\phi}^{a}_{n}
(\equiv \phi^{a}/\phi)$ such as
\be \label{angle}
\hat{\phi}^{a}_{n}=(\sin n\theta\cos n\varphi,\,\sin n\theta\sin n\varphi
,\, \cos n\theta ),
\ee
where the allowed $n$ is 0 or 1 in order to render the scalar amplitude $\phi$
a function of $r$ only.
Under this assumption, the Euler-Lagrange equation becomes
\be \label{equation}
\frac{d^{2}\phi}{dr^{2}}+\frac{2}{r}\frac{d\phi}{dr}
-\delta_{n1}\frac{2}{r^{2}}\phi =\frac{dV}{d\phi},
\ee
where $\delta_{n1}$ in the third term of Eq.(\ref{equation}) denotes Kronecker
delta for $n=0$ or 1.

The condition that the theory should be in
false vacuum at spatial infinity fixes the boundary value of
field, {\it i.e.} $\lim_{r\rightarrow\infty}\phi\rightarrow 0$.
To be nonsingular solution at the origin of coordinates, the boundary
condition is
\be
\left\{ \begin{array}{ll}
\displaystyle \left.\frac{d\phi}{dr}\right|_{r=0}=0 & \mbox{if $n=0$}\\
\displaystyle \phi(0)=0 & \mbox{if $n=1$.}
\end{array}\right.
\ee
When $n=0$, it is well-known bubble solution at finite temperature
\cite{Lin}. We will
analyze $n=1$ case and show that there always be $n=1$ solution
if the equation contains $n=0$ solution. A brief argument of the existence
of $n=1$
solution is as follows.

If we regard the radius $r$ as time and the scalar amplitude $\phi (r)$
as the coordinate of a particle, Eq.(\ref{equation}) describes a
one-dimensional motion of a unit-mass hypothetical particle under
the conserved force due to the potential
$-V(\phi)$ and two
nonconservative forces, {\it i.e.} one is the friction of
time-dependent coefficient $-\frac{2}{r}\frac{d\phi}{dr}$
and the other is
time-dependent repulsion $\frac{2}{r^{2}}\phi$. Hence,
in the terminology of Newton equation, $n=1$ solution in Fig. 1 is interpreted
as the motion of a particle that starts at time zero at the origin
$(\phi(0)=0)$, turns at an appropriate nonzero position at a certain
time $(\phi(t_{turn})=\phi_{turn})$, and stops at the
origin at infinite time $(\phi(\infty)=0)$.

At first, if one considers a hypothetical particle at the origin
at time zero, it is
accelerated by the time-dependent repulsion of which coefficient is
divergently large for small $r$.  Since one
initial condition is fixed by the starting point $(\phi(0)=0)$,
such motions near the origin are characterized by another initial
condition $C$
\be\label{phi0}
\phi(r)\approx C\left\{ r+{\cal O}(r^{3})+\cdots\right\},
\ee
where $C$ is the initial velocity of a particle which should be tuned by the
proper boundary condition at
infinite time $(\phi(\infty)=0)$. From now on let us consider a set
of solutions specified by a real parameter $C$ and show that there
always exists the unique motion of $\phi(\infty)=0$ for an appropriate
$C$. When $C$ is sufficiently large, the acceleration near the origin due to
time-dependent repulsion and conservative force is too strong and then the
particle overshoots the top of potential $-V(\phi=v)$ and goes to infinity
$(\phi(\infty)=\infty)$, despite the deceleration due to the friction.
Since the solution of our interest is the motion which includes a
return, we will look
at the motions of which $C$ is smaller than a critical value $C_{top}$.
Here $C_{top}$ gives
the motion that the particle stops at the hilltop of the potential
$-V$ at infinite time $\phi(\infty)=v$.
Suppose that $C$ be too small, {\it i.e.} $C$ is smaller than another
critical value $C_{0}\; (C_{0} <C_{top})$, particle turns at a point smaller
than
$\phi_{0}$ ($0<\phi_{0}<v$) where $-V(\phi=\phi_{0})=0$ and thereby it can not
return to the origin. Moreover, the
particle which turns at a point too close to $v$ arrives at the origin
at a finite time, since the time-dependent friction and repulsion during
returning do not play a role due to the vanishing of their coefficients
$1/r$ and $1/r^{2}$ while the particle stays near $v$ for sufficiently
long time.
By continuity for an appropriate $C\;(C_{0}<C<C_{top})$ there is a solution
which
describes the motion that starts at time zero at the origin turns at a position
$\phi_{turn}$ between $\phi_{0}$ and $v$, and then stops at the origin
at infinite time.
Here for the existence proof of $n=1$ solutions, we have used the properties
of scalar potential $V(\phi)$ that it has a local minimum at $\phi=0$,
a local maximum at $\phi_{bottom}$ ($0<\phi_{bottom}<\phi_{0}$) and an
absolute minimum at $v$.
It is exactly the same condition as that for the existence of $n=0$ solution.
Hence we completes our argument that the equation always contains $n=1$
solutions
when it has $n=0$ solution. A rigorous proof for the existence of $n=1$
solution is demonstrated in ref.\cite{KKK}.
We have no analytic form of solutions, so numerical solutions for $n=0$ and
$n=1$ bubbles are given in Fig. 1.

Suppose that each classical solution describes a critical
bubble corresponding to a different decay channel,
from the profile of energy density $T^{0}_{\;0}$ (see Fig. 2) we can read
the following characteristics of
$n=1$ bubbles. First, the boundary value of
scalar field at the origin always has that of false vacuum, {\it i.e.}
$\phi(0)=0$, so this
implies that there remains a false-vacuum core inside the true-vacuum region of
the bubble
due to the winding between $O(3)$ internal symmetry and spatial rotation.
Second, the value of energy density has a local minimum at the origin,
{\it i.e.} $3C^{2}/2+V(0)$ in which $C$ is a constant appeared in
Eq.(\ref{phi0}), increases to the maximum at $R_{m}$ in Fig. 2 and decays below
$V(0)$.
This implies that a matter droplet is created inside the bubble and is
surrounded by inner bubble wall with size of order $R_{m}\sim 1/m_{Higgs}
=\sqrt{4(3+2\alpha)\lambda}v$.
Third, if we read the expression of energy
density when the scalar amplitude has maximum value $\phi=\phi_{turn}$,
its derivative term vanishes and then it becomes
\be\label{turn}
T^{0}_{\;0}=\frac{\phi_{turn}^{2}}{r^{2}}+V(\phi_{turn}).
\ee
The potential term $V(\phi_{turn})$ can be neglected in thin-wall limit,
so the object inside the bubble has a long-range hair which penetrates the
inner
bubble wall.
Since the central region of this matter is in false symmetric vacuum due to
the hedgehog ansatz of scalar field in Eq.(\ref{angle}) and
for large $r$ the scalar field goes to the true broken vacuum but the
long-range tail of energy proportional to $r$ has a cutoff at the outer bubble
wall at $R_{n=1}$ in Fig. 2, we can interpret the matter aggregate inside the
$n=1$
bubble as a global monopole of size $R_{m}$ \cite{Vil}. Even though there is
a no-go theorem for static scalar objects in spacetime
dimensions more than two \cite{Der}, the global monopole of this case can
be supported inside the Euclidean bubble configuration as a smooth
finite-energy configuration due to a natural cutoff introduced by the outer
bubble wall. Fig. 2 shows that the radius of $n=1$ bubble is
larger than that of $n=0$ bubble. It can be easily understood by the
conservation
of energy, {\it i.e.} the additional energy used to make a matter aggregate is
equal to
the loss of energy due to the increase of the radius of bubble.

Though we do not have any analytic solution, we can estimate $B_{n}$ in
Eq.(\ref{decay}) by use of the obtained bubble configurations through numerical
analysis for given parameters of theory. It has already been proved that
$n=0$ solution describes the nontrivial solution of lowest action \cite{CGM}
and from Fig. 3 we read that the value of action for $n=1$ solution,
$vB^{'}_{1}/T$, is larger than that of $n=0$, $vB^{'}_{0}/T$, irrespective of
the
shape of scalar potential. Fig. 3 also shows that the ratio
$B^{'}_{1}/B^{'}_{0}$ becomes small while the difference $B^{'}_{1}-
B^{'}_{0}$ increases as the size of bubble becomes large in comparison with
the mass scale of theory, {\it i.e.} thin-wall limit. This can be understood
as follows; the amount of energy consumed to support a global monopole at
the center of bubble is proportional to the
radius of bubble due to its long-range tail (see Eq.(\ref{turn})), however
the free energy to support the bubble itself, $vB^{'}_{0}$ or $vB^{'}_{1}$,
is proportional to the cubic of bubble radius in thin-wall limit.

The next task is to estimate the pre-exponential factor $A_{n}$ in
Eq.(\ref{decay}). The scheme of computing $A_{0}$ for $n=0$ bubble solution
which is the lowest action solution with one negative mode was given in
the second paper of Ref.\cite{Col1}. Here, let us attempt to calculate the
pre-exponential factor $A_{1}$ for $n=1$ bubble. Considering the small
fluctuation around $n=1$ bubble $\delta\phi^{a}_{n=1}=\sum_{k}c^{a}_{k}
\psi^{a}_{k}$, we obtain a Schr\"{o}dinger-type equation for three particles
in three dimensions
\be\label{sch}
\biggl(-\nabla^{2}\delta^{ab}+\hat{r}^{a}\hat{r}^{b}\frac{d^{2}V}{d\phi^{2}}
\biggr|_{\phi^{a}_{n=1}}+(\delta^{ab}-\hat{r}^{a}\hat{r}^{b})\frac{1}{\phi}
\frac{dV}{d\phi}\biggr|_{\phi^{a}_{n=1}}\biggr)\psi^{b}_{k}=\lambda_{k}
\psi^{a}_{k}.
\ee
It looks too difficult to solve this equation since it includes $\theta$ and
$\varphi$ dependent off-diagonal terms in its Hamiltonian and the potential
form is only determined numerically, however the eigenfunctions of six
zero modes due to three translations and three rotations are explicitly given
in a form
\bea\label{zero}
\psi^{a}_{k,\, i}&=&N_{t}\nabla_{i}\phi^{a}_{n=1}\nonumber\\
&=&N_{t}\biggl\{\hat{r}_{i}\hat{r}^{a}
\frac{d\phi_{n=1}}{dr}+(\delta_{i}^{\;a}-\hat{r}_{i}\hat{r}^{a})
\frac{\phi_{n=1}}{r}\biggr\},
\eea
and
\be\label{azero}
\psi^{a}_{k,\, i}=N_{r}\epsilon_{ijk}x^{j}\nabla^{k}\phi^{a}_{n=1}
=N_{r}\epsilon_{ija}\hat{r}^{j}\phi_{n=1},
\ee
where $N_{t}$ and $N_{r}$ are normalization constants.
Here we give few comments on the number of negative modes. First, every
component of translational zero-mode eigenfunction in
Eq.(\ref{zero}) has
$(\theta,\varphi)$-dependence where its radial part,
$\frac{d\phi_{n=1}}{dr}-\frac{\phi_{n=1}}{r}$, has single node at the origin,
and each $a=i$ component includes an additional
$(\theta,\varphi)$-independent part, $\frac{\phi_{n=1}}{r}$, which has no node.
Every $a\neq i$ component of rotational ones, $\phi_{n=1}$, has one node at the
origin. For the sake of
simplicity, let us examine the problem for the perturbation
with specific direction; $(i)$ one for amplitude ($\delta\phi^{a}_{n=1}=
\hat{r}^{a}\delta\phi_{n=1})$ and $(ii)$ two for transverse
($\delta\phi^{a}_{n=1}=(\delta^{ab}-\hat{r}^{a}\hat{r}^{b})
\delta\phi^{b}_{n=1}$). Under these fluctuations Eq.(\ref{sch}) reduces to
\be\label{nonz}
(-\nabla^{2}+U(r))\psi^{a}_{k}=\lambda_{k}\psi^{a}_{k},
\ee
where $U(r)=\left.\frac{d^{2}V}{d\phi^{2}}\right|_{\phi_{n=1}}$ for
the first case $(i)$ and $U(r)=\left.\frac{1}{\phi}\frac{dV}{d\phi}
\right|_{\phi_{n=1}}$ for the second case $(ii)$. When $
U(r)=\left.\frac{d^{2}V}{d\phi^{2}}\right|_{\phi_{n=0}}$, each single `$a$'
component of
Eq.(\ref{nonz}) is nothing but the fluctuation equation for $n=0$ bubble
which contains a nodless $s$-wave mode as the unique negative mode
\cite{Col1,SCol}. For the fluctuation of scalar amplitude for $n=1$ bubble,
the lowest mode can not be nodless $s$-wave mode but $l=1$ mode with single
node
at $r=0$ since $\psi^{a}_{k}$ should have $\theta$ and $\varphi$ dependence
proportional to $\hat{r}^{a}$ even though the operator in Eq.(\ref{nonz})
is a scalar operator. It is analogous for the perturbation to the transversal
directions. It implies that the eigenfunction of negative mode may take a
somewhat complicated form which depends on angels. Second, the $n=1$ solution
which is a parity-odd bounce solution, $\phi(r=0)=\phi(r=\infty)$, is
supported by assuming the winding between three spatial rotations and those
in internal space. It seems that the argument for the system of quantum
mechanics with one time variable in Ref.\cite{SCol} does not forbid directly
the existence of $n=1$ solution as a bubble
configuration. We know very little on the counting of negative modes,
and then there should be further work on this issue.
Since the operator in Eq.(\ref{sch}) is even under parity transformation and
is covariant under the rotation, {\it i.e.} $x^{i}\rightarrow O^{ij}x^{j}$ and
$\psi^{a}_{k}(x^{'})=O^{ab}\psi^{b}_{k}(x)$ where both $O^{ij}$ and $O^{ab}$
are the elements of $O(3)$ group, vector spherical harmonics is a method to
investigate the modes of which the eigenfunctions can be chosen to be
either even or odd in parity \cite{Tam}.

In order to compute the decay rate accurately, we should calculate the nonzero
modes $\lambda_{k}$, which is extremely difficult even for $n=0$ bubbles.
However, the dimensional estimate may reach a rough result such as \cite{Lin}
\be
\frac{\Gamma^{(1)}}{\Gamma^{(0)}}\sim\biggl(\frac{B^{'}_{1}}{B_{0}^{'}}
\biggr)^{\frac{6}{2}}\exp\Bigl[-\frac{v}{T}(B^{'}_{1}-B^{'}_{0})\Bigr].
\ee
At high temperature limit, $T>>v$,
both decay rates for $n=0$ and 1 bubbles, of course, increase and, moreover,
the relative decay rate of $n=1$ bubble to that of $n=0$ is also enhanced
exponentially. When $v/T\sim 10^{-1}$, the
order of $\Gamma^{(1)}/\Gamma^{(0)}$ is around 0.17 in a thin-wall case
($\lambda=1$, $\alpha=0.12$)
and it is around 28 in a thick-wall case ($\lambda=1$, $\alpha=0.47$).
Therefore, at high temperatures and in thick-wall case, this $n=1$ bubbles
can be preferred to those of $n=0$, and obviously
the existence of another decay channel can enhance considerably the total
nucleation rate of bubbles, $\Gamma=\Gamma^{(0)}+\Gamma^{(1)}$, except for
the bubbles with extremely thin wall.

The main consideration of the paper was to gain an understanding as to how
bubbles with global monopoles nucleated in first-order phase transitions at
high temperatures and the next question will be how the high-temperature
bubbles grow \cite{Ste}, particularly at the site of global monopole. Once we
assume
these bubbles in early universe where the gravity effect should be included,
the bubbles
with solitons may result in interesting phenomena \cite{KMS} in relation with
inflationary models, {\it i.e.} the inflation in the cores of topological
defects \cite{Vil2}.

The author would like to express deep gratitude to Jooyoo Hong, K. Ishikawa,
Chanju Kim, Kimyeong Lee, K. Maeda, V. P. Nair, Q. Park, N. Sakai, A. I. Sanda,
S. Tanimura, E. J. Weinberg, Y. S. Wu and K. Yamawaki for valuable discussions
and also thanks Kyung Hee University and Hanyang University for their
hospitality. This work is supported by JSPS under $\#$93033.

\def\hebibliography#1{\begin{center}\subsection*{References
}\end{center}\list
{[\arabic{enumi}]}{\settowidth\labelwidth{[#1]}
\leftmargin\labelwidth	  \advance\leftmargin\labelsep
    \usecounter{enumi}}
    \def\newblock{\hskip .11em plus .33em minus .07em}
    \sloppy\clubpenalty4000\widowpenalty4000
    \sfcode`\.=1000\relax}

\let\endhebibliography=\endlist

\begin{hebibliography}{100}
\item[$\ast$] E-mail address: yoonbai$@$eken.phys.nagoya-u.ac.jp
\bibitem{Lan} J. S. Langer, Ann. Phys. {\bf 41}, 108 (1967).
\bibitem{Col1} S. Coleman, Phys. Rev {\bf D15}, 2929 (1977); C. Callan and S.
Coleman,
{\it ibid} {\bf D16}, 1762 (1977).
\bibitem{Lin} A. D. Linde, Phys. Lett. {\bf B70}, 306 (1977); {\it ibid}
{\bf B100}, 37 (1981); Nucl. Phys. {\bf B216}, 421 (1983).
\bibitem{KT} I. Affleck, Phys. Rev. Lett. {\bf 46}, 306 (1981);
E. W. Kolb, and I. I. Tkachev, Phys. Rev. {\bf D46}, 4235 (1992); S. D. H. Hsu,
Phys. Lett. {\bf B294}, 77 (1992).
\bibitem{KKK} C. Kim, S. Kim and Y. Kim, Phys. Rev. {\bf D47}, 5434 (1993).
\bibitem{Vil} M. Barriola and A. Vilenkin, Phys. Rev. Lett. {\bf 63}, 341
(1989).
\bibitem{Der} G. H. Derrick, J. Math. Phys. {\bf 5}, 1252 (1964); R. Hobart,
Proc. Phys. Soc. {\bf 82}, 201 (1963).
\bibitem{CGM} S. Coleman, V. Glaser and A. Martin, Comm. Math. Phys. {\bf 58},
211 (1978).
\bibitem{SCol} S. Coleman, Nucl. Phys. {\bf B298}, 178 (1988).
\bibitem{Tam} I. Tamm, Z. Phys. {\bf 71}, 141 (1931); E. Weinberg, Phys. Rev.
{\bf D49}, 1086 (1994); Y. Kim, in preparation.
\bibitem{Ste} P. J. Steinhardt, Phys. Rev. {D25}, 2074 (1982).
\bibitem{KMS} Y. Kim, K. Maeda and N. Sakai, in preparation.
\bibitem{Vil2} A. Vilenkin, Phys. Rev. Lett. {\bf 72}, 3137 (1994); A. Linde
and
D. Linde, Stanford university preprint SU-ITP-94-3; A. Linde, Phys. Lett.
{\bf B327}, 208 (1994).
\end{hebibliography}

\newpage

\begin{center}\section*{\large\bf Figure Captions}\end{center}
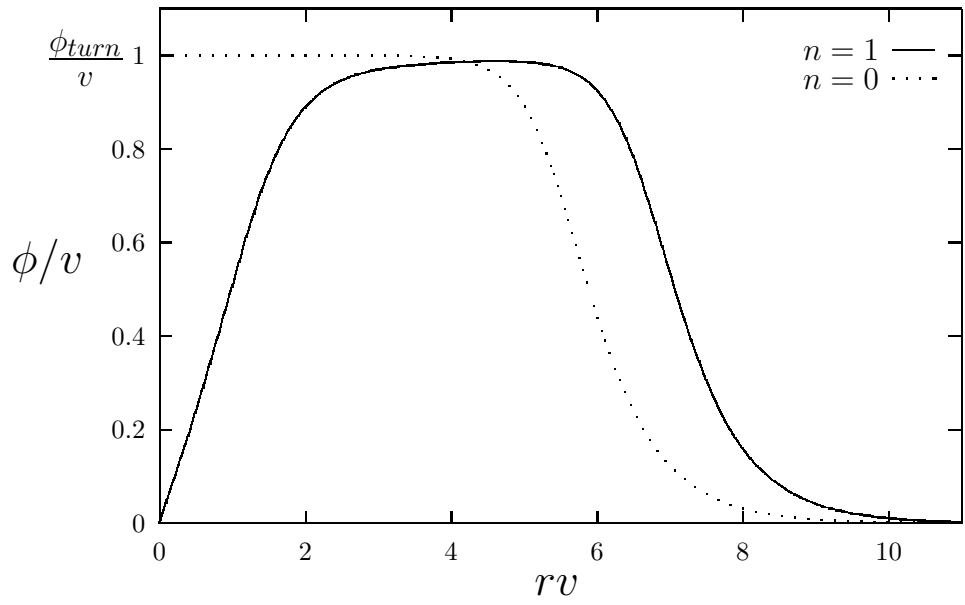
\noindent FIG. 1. Plot of bubble solutions. $n=0$
and $n=1$ configurations are shown as dotted and solid lines,
respectively.
The parameters chosen in the figures are: $\lambda=1$,
$\alpha=0.12$, and $V_{0}=-0.12v^{4}$.\newline

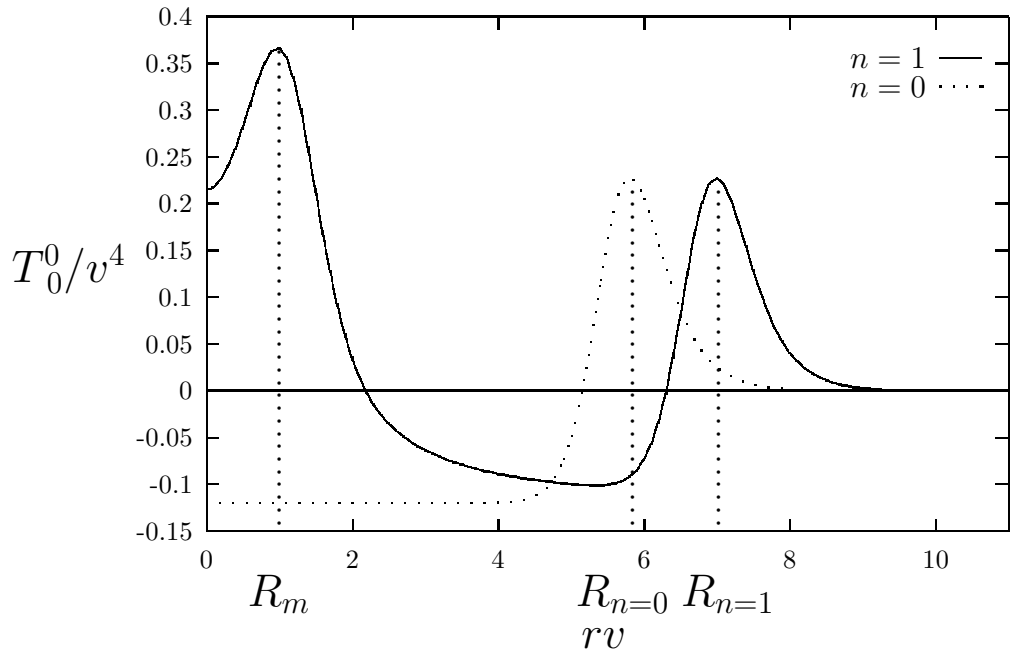
\noindent FIG. 2. Plot of energy density $T^{0}_{0}$.
The parameters chosen in the figures are: $\lambda=1$,
$\alpha=0.12$, and $V_{0}=-0.12v^{4}$ which is the minimum of energy
density.\newline

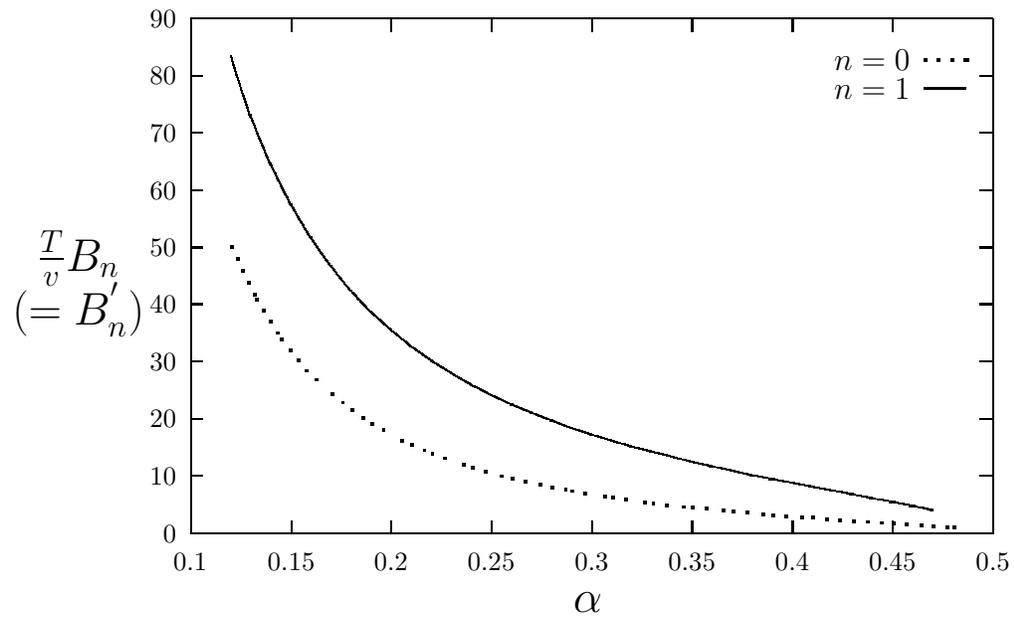
\noindent FIG. 3. Plot of action $S_{E}$ (or equivalently $B_{n}$) as a
function of
$\alpha$. Another parameters are chosen as $\lambda=1$ and
$V_{0}=-\lambda\alpha v^{4}$.

\newpage

\setlength{\unitlength}{0.240900pt}
\ifx\plotpoint\undefined\newsavebox{\plotpoint}\fi
\sbox{\plotpoint}{\rule[-0.200pt]{0.400pt}{0.400pt}}%


\end{document}